# In-Depth Search for a Coupling between Gravity and Electromagnetism with Steady Fields

M. Tajmar[1], M. Kößling and O. Neunzig

TUD Dresden University of Technology, Institute of Aerospace Engineering, 01307 Dresden, Germany

**Abstract**

Any means to control gravity like electromagnetism is currently out of reach by many orders of magnitude even under extreme laboratory conditions. Some often poorly executed experiments or pseudoscience theories appear from time to time claiming for example anomalous forces from capacitors that suggest a connection between the two fields. We developed novel and high resolution horizontal-, vertical- and rotation-balances that allow to test electric devices completely shielded and remotely controlled under high vacuum conditions to perform the first in-depth search for such a coupling using steady fields. Our testing included a variety of capacitors of different shapes and compositions as well as for the first-time solenoids and tunneling currents from Zener diodes and varistors. A comprehensive coupling-scheme table was used to test almost all combinations including capacitors and solenoids with permittivity and permeability gradients as well as capacitors and varistors within crossed magnetic fields. We also tested a crossed-coil producing helical magnetic field lines as well as interactions between a pair of shielded toroidal coils to look for proposed extensions to Maxwell's equations. No anomalous forces or torques down to the Nanonewton or Nanonewton-Meter range were found providing new limits many orders of magnitude below previous assessments ruling out claims or theories and providing a basis for future research on the topic.

## 1. Introduction

Gravity is the weakest of the four fundamental forces, yet it appears as the dominant limit that keeps us on the ground making space travel hard and interstellar travel impossible [1] with no chance to manipulate or escape its always attractive property. On the other hand, electromagnetism, the second non-nuclear fundamental force, is the basis for our modern age thanks to the countless possibilities of changing, shielding or amplifying electromagnetic fields. However, according to general relativity theory, neutron star mass densities would be required to do the same with gravity in the laboratory [2], which is far out of reach.

It is quite probable that every reader of this paper was wondering more than once if there is any possibility to manipulate gravity like electromagnetism. Without doubt, such a discovery would have significant consequences leading to new physics and technologies that are presently only science fiction. This thought can be even traced back to one of the inventors of electromagnetism, Michael Faraday, who performed experiments as early as 1849 looking for any connection without success [3].

Some hundred years ago, a patent appeared claiming that a high-voltage capacitor will move towards its positive pole suggesting a deeper link between gravity and electromagnetism, a phenomenon that became later well known as the Biefeld-Brown effect [4]. Several follow-up patents were published by Brown until the 1960s identifying asymmetrically shaped-capacitors as the most promising geometry, popular science magazine articles [5], books as well as scientific publications, mostly of low quality, appear until the present day claiming that this effect is real and of gravitational origin [6].

[1] Email: martin.tajmar@tu-dresden.de

Mainstream science quickly identified the capacitor movement as a phenomenon which is called Electrohydrodynamics (EHD) or Corona-wind, where a discharge between the electrodes collides with neutral gas molecules, transferring momentum which creates thrust [7], [8]. Any anomalous force that may still be present but hidden due to the large background corona wind force was investigated by putting the whole capacitor assembly in a closed box at the end of a vertical pendulum. No force was found within an error bar of approximately 10 µN applying 40 kV and 0.6 mA discharges [9]. Recently, we looked again for force or weight changes for high-voltage capacitors of various shapes, focusing on solid dielectrics and no discharge following Brown's original experiments, using an off-the-shelve precision balance. The test samples were mounted again in a closed box with thermal and electrostatic shielding as well as a battery-powered on-board high-voltage generator in order to reduce buoyancy or electromagnetic interactions as much as possible. Again, no anomalous forces were found down to around 3 µN ruling out not only other experimental claims but almost all of the proposed theories [10].

With anomalous we mean anything that is larger than $E=m.c^2$, which shows that electromagnetic energy is a contributor to mass and hence establishes a classic gravity-electromagnetism link. However, $c^2$ is a huge number requiring energies for measurable mass changes in the laboratory that are nearly impossible. For example, an off-the-shelve supercapacitor with a charging voltage of 3 V and a capacity of 500 F weighs 96 g. By charging it up, we would classically expect a change in mass of about $2.5\times10^{-11}$ g or $2.5\times10^{-7}$ µN, well outside of available measurement ranges.

Of course, a connection may still exist but at a smaller value as all tests up to now could only rule out forces down to the µN range, which is accessible to most well-equipped university laboratories. There is no doubt that anomalous forces down to such values would therefore already be well know if they exist. In addition, other coupling schemes might be present apart from a simple capacitor.

We therefore decided to perform an in-depth search of possible couplings between gravity and electromagnetism in a systematic manner to test almost all combinations with the best available balance technology pushing our measurement boundary by many orders of magnitude to the nano-Newton range. Here we will answer such basic questions like: Does a capacitor or solenoid produce a self-force or does the weight of electrically or magnetically polarized substances change? This has not been rigorously assessed in the scientific literature up to now. Our tests included capacitors and solenoids with different shapes, core materials and also a novel crossed-coil configuration in addition to Zener diode/varistor arrays.

Previous research on falsifying claims on potentially revolutionary propellant-less thrusters led us to the development of balances, that appear ideally suited for such an investigation taking numerous lessons-learned into account to eliminate buoyancy effects, center of gravity shifts or electromagnetic environment interactions [11], [12], [13]. Our latest balances feature:

- Operation in high vacuum
- Use on-board battery power and high-voltage/high-current power supplies
- Wireless data acquisition and command
- Multi-layer shielding using highly-magnetic permeable material

We used our balances to test for linear forces, weight changes as well as torques to get the first thorough picture of possible gravity-electromagnetic couplings and establish boundaries against which theories and other claims can be reliably tested.

**Gravity-Electromagnetic Coupling Schemes**

Figure 1 illustrates our systematic investigation with gravitational parameters like gravitational field and mass on the vertical axis and electromagnetic parameters used in Maxwell's equations on the horizontal axis. We also included an additional row to allow for non-Maxwellian parameters that are up to now speculative and unconfirmed. A crossed cell marks an incompatibility between vector and scalar quantities such as coupling between a charge and a gravitational field or a current and a mass. Green areas are assessed in the present work, which is a good part of all possible coupling combinations. If a gravitational field is present, according to the equivalence principle, it is indistinguishable from a force acting on our test article. Hence, our balances should be able to pick up any anomalous gravitational field. Mass changes are measured with the beam balance by looking for variation in weight.

The coupling table includes an incomplete number of theories or experimental claims that would fit in the box and for some of them experimental results are already available. The results in this paper provide an experimental basis with either first-time measurements or much higher resolution than previous work and put boundaries on all boxes. It's clear that our work is non-exhaustive and we had to limit ourselves to experiments with stationary conditions (stationary E and B fields, DC currents, etc.). This needs to be extended in the future (first results are presented here [14]). Also scale and geometry may play a role that needs to be addressed in a follow-up assessment.

Here we will briefly discuss each combination:

**Electric Field *E*:** Since both electric and the gravitational fields are vectors and the fact that both can originate from stationary sources like charge and mass, a possible link between them is the most obvious and possibly best investigated one. It leads back to the Biefeld-Brown effect and the question, if a capacitor generates a force field in addition to its electric field. The most representative theory for such a link was proposed by Ivanov, suggesting that static electric or magnetic fields induce Weyl-Majumdar-Papapetrou solutions for the metric of spacetime [15], [16]. This shall create a gravitational acceleration some 22 orders of magnitude stronger than classically expected. The predicted force on a simple parallel-plate capacitor is given by

$$F = \sqrt{G\varepsilon_0\varepsilon_r}\frac{m}{d}V = \sqrt{G\varepsilon_0\varepsilon_r}\rho AV \tag{1}$$

where $G$ is Newton's gravitational constant, $\varepsilon_0$ and $\varepsilon_r$ the electric constant and electric permittivity respectively, $m$ the dielectric's mass, $d$ its thickness, $\rho$ its density, $A$ its area and $V$ the applied voltage. For high-permittivities (high-k), Ivanov's root gravity theory suggests milli-gram weight changes for classical off-the-shelve capacitors that were ruled out by our earlier experiments (as well as a number of other theories and claims not discussed here) [10], however, for low-permittivities (low-k) such as PTFE dielectrics, a weight change was predicted, which was below our previous resolution.

In addition to force, electric fields could also induce a mass change due to electric polarization of the dielectric. Assis proposed an extension to Weber electrodynamics (an early attempt to explain electromagnetism with a velocity and acceleration dependent Coulomb force) where higher-order terms lead to always attractive gravity-like forces from oscillating electric dipole-dipole interactions [17], [18]. Although the mass model had too many free parameters to derive a particle's mass directly, it showed interesting insights into a possible link between gravity, electromagnetism and even quantum behavior [19], [20]. Alignment of electric dipoles could therefore lead to changes in the sample's mass. Our earlier assessment showed no weight changes for regular capacitors within our resolution [10].

**Magnetic Field *B*:** Similar to the electric case, a magnetic field may create a gravitational field. However, as magnetic field lines are rotational in nature, here we expect gravitational-like fields that lead to torques rather than linear forces. Surprisingly, this coupling scheme had gotten much less attention compared to the capacitor case with no comparison to experimental data according to our knowledge. Ivanov's root gravity theory predicts a similar force law in this case given by

$$F = \sqrt{G\mu_0\mu_r}\rho VH \tag{2}$$

where $\mu_0$ and $\mu_r$ are the magnetic free-space and relative permittivity, and $H$ is the magnetic field strength and $V$ the volume. A toroidal coil geometry seems perfect to test for torques and to minimize any stray fields that can interact with the environment. This leads to a force acting in the middle of the solenoid as

$$F = \sqrt{G\mu_0\mu_r}\frac{\rho VNI}{2\pi r} = \sqrt{G\mu_0\mu_r} \cdot \rho ANI \tag{3}$$

where $N$ is the number of turns, $I$ the current and $r$ the average radius and $A$ the cross section of the toroidal coil. It will then generate a torque along its symmetry axis given by

$$\tau = \sqrt{G\mu_0\mu_r} \cdot \rho ArNI \tag{4}$$

Similar to the electric case, we also want to test if simple magnetic polarization of various substances can influence mass. As a special case, Alzofon [21] predicted that changes in nuclear spin-polarization lead to mass changes, which was recently investigated by us in a dedicated experiment with a zero result [22].

**Crossed Electric and Magnetic Fields (*E×B*):** The cross product of two vectors is still a vector and we can ask the same question if there is any anomalous coupling to the gravitational field. This cross product is of course closely related to the Poynting vector, describing the energy transfer of an electromagnetic wave. Radiative energy is a well-known method to produce thrust like the photon force from a laser. There have been claims that a microwave inside a conical-shaped resonator produces an anomalous force (EmDrive), which was falsified by us tracing it back to a setup-related effect [11].

*E×B* fields have been proposed to trigger vacuum or Casimir forces [23], [24] that extract momentum from the vacuum, similar to an anomalous force, which may be detectable with our setup. So far, no experimental values have been reported and numerous critics on the concept have been published (e.g. [25]). We tested *ExB* effects as well as possible mass changes with both capacitors and varistors.

**Charge *q*:** As a scalar quantity, charge may share a deeper connection to mass. Assis predicted that based on Weber electrodynamics, the inertial mass of a moving charge should vary due to the presence of surrounding stationary charges [26]. This can be understood as a contribution of electrostatic potential energy to mass. However, it is not possible to assign this mass change to individual charges but only to the overall system of interacting charges as recently demonstrated by us experimentally [27].

Another approach is closely linked to the Kaluza-Klein theory, which provides an elegant link between gravity and electromagnetism in 5 dimensions. Here, an electro-gravity buoyancy force for electrically-charged objects is predicted in the Earth's environment due to an additional long-range scalar field. No such force was detected experimentally looking for weight changes from large unipolar charged up masses [28]. Also claimed mass changes detected with charged-up torsion pendulums (Saxl effect) were later revealed to be simple electrostatic interactions that were not properly shielded [29].

**Current *I*:** As the current flows parallel to the electric field, this case is related to our first coupling scheme outlined above with the addition that there are now moving charges. Furthermore, it might be of interest to look for currents in a special environment, which may trigger an anomalous force. For example, we recently investigated if superconducting currents trigger such forces using both YBCO and BSCCO samples achieving a null result within our resolution [30]. Another source may be discharge currents with a divergent current flow, proposed to trigger an additional mass flow in a 5D extension to Maxwell's equations [31], like those in classical asymmetrically shaped Biefeld-Brown experiments or capacitors with leakage currents. Both cases were already experimentally assessed as described above [9], [10].

However, there exists another current type that may be of interest: tunneling currents. No data has been published to our knowledge on force or mass changes associated with quantum tunneling.

**Speed of Light *c* and Vacuum Properties $\varepsilon_0, \eta_0$:** The speed of light and vacuum properties are usually fixed constants that appear in equations. Obviously, changing fundamental constants will have an effect on gravity. The only known method is vacuum engineering using Casimir cavities. For example, the speed of light perpendicular to Casimir plates is predicted to be larger than parallel to the plates [32], however, the amount is miniscule and un-measurable. Vacuum/Zero-point fields have also been used to explain aspects of inertia, however not with much success [33]. Another point of view is that vacuum is a type of superfluid and fundamental constants may change above/below a Curie-type temperature. The weight of samples has been investigated by our team in both the cryo- and high-temperature regime without any measurable difference [34], [35].

Our present work does not investigate these type of coupling aspects as it involves very different setups.

**Non-Maxwell:** This row includes any deviation from standard Maxwellian electromagnetism that could lead to gravity-coupling. There have been many modifications proposed such as consequences from a non-zero photon-mass (Proca equations), Born-Infeld non-linear electromagnetics which tackle the problem of infinite self-energies and point charges, additions due to magnetic monopoles, more dimensions or scalar fields. Most of them have no real-world consequences in the laboratory. An exception is the modified Kaluza-Klein theory proposed by Mbelek and Lachièze-Rey, which predicts that solenoids will generate a new force based on a scalar field, that cannot be shielded like traditional magnetic fields [36]. Mbelek claims to have even measured this effect in the laboratory [37], [38], however, a recent experimental assessment by our group show that the origin of the effect are most likely diamagnetic forces [39]. However, no real systematic assessment of this effect has been done up to now like the interaction of two shielded solenoids.

## 2. Experimental Setup

### 2.1 Overview

Here we will give an overview of the overall experimental setup and the general execution of the experiments as well as the common measurement technology. All tests were done inside a large cylindrically shaped vacuum chamber with a diameter of 0.9 m and a length of 1.5 m. In order to suppress vibrations from the environment, the chamber is put on top of an optical table that is mounted on a large concrete block, which is de-coupled from our building, in our basement laboratory. A roughing (Edwards XDS35i) and turbo-pump (Pfeiffer Turbo HiPace 2300) enables operation in high vacuum usually in the $10^{-7}$ mbar range. An illustration of the overall setup is shown in Fig. 2

All balances use an attocube IDS3010 laser interferometer to read out its position and a voice-coil in combination with a Keithley 2450 SourceMeter to accurately calibrate the balance with a known force to calculate the displacement-force relationship. A LabView software is used to control both balance and experiment as well as to monitor environmental conditions such as vacuum pressure and in some cases temperature or magnetic fields. A script language was implemented to ensure the same execution of all tasks and to repeat measurements in a very reproducible manner. On-board battery power and (multi) mu-metal-shields provide very effective electromagnetic shielding from the environment, which is an important error source for these kinds of experiments. Charging cables for the battery as well as signal wires are connected through the balance to the measurement box using specially-designed liquid-metal feedthroughs. Charging is only done prior to the experiment in order not to disturb the measurement. For low voltages (solenoids and low-voltage varistors), a DPS/DPH miniature power supply module was used that could be remotely commanded using Bluetooth. High voltages (capacitors and high-voltage varistors) were generated using either off-the-shelve DC-DC converters (like an EMCO CB101, Spellman UM40P30 or ISEG 4010512) or a custom-built board for 40 kV designed to work in high vacuum. An ESP32 data acquisition board was used to command the high voltage converters and monitor output voltages and currents as well as temperatures.

High vacuum and monitoring of possible outgassing (e.g. due to high currents through our test articles) are important to remove all gas-related effects like buoyancy.

The general test execution was done as follows:

1. Mount test article on balance, pump down chamber and wait until nighttime for optimal noise and resolution.
2. Perform calibration with voice-coil.
3. Do tests like charging up capacitor or solenoid and repeat each measurement for every voltage or current tens of times (typically more than 80). This increases statistical significance and is used to reduce the noise by using signal averaging of all profiles with the same condition. Each measurement is executed by following down-time, ramping up, constant voltage/current, ramping down and another down-time sector with defined time steps. Post-processing is used to remove thermal-drifts as well as to identify and remove outliners (e.g. large balance offsets due to a truck driving on the road next to our laboratory). A detailed discussion on our signal processing can be found here [11].
4. Perform another calibration to check that the balance did not change its spring constant.
5. Dismount and prepare for next measurement.

### 2.2 Balances

This chapter will give a brief overview of all balances used, more details on our balances can be found here [11], [40].

#### 2.2.1 Inverted Counterbalanced Double-Pendulum Balance (Horizontal Measurement)

A double-pendulum balance (55 x 74 x 70 cm³) consists of two platforms that host the experiment on the top and a counter-weight on the bottom. As illustrated in Fig. 3, a tripod then connects the two platforms with a total of nine frictionless flexural pivots. Such a pendulum has significant advantages for measuring linear vertical forces as the platform-concept greatly reduces sensitivity to mass shifts. In addition, the test article is hanging down from a pivot in the middle of the upper platform which proved to be very effective to measure small forces with minimum drifts. Up to three mu-metal boxes were assembled around the test article to provide high magnetic shielding, featuring an experiment volume of approximately 30.6 x 26 x 20 cm³. Three shields were used for solenoid measurements, capacitor or diodes were tested with only one shield as the magnetic field from these devices was small.

In order to test for a Mbelek-like scalar force originating from a solenoid but not shielded by the mu-metal boxes, we also tested one toroidal coil in our measurement box and another toroidal coil inside a similar three mu-metal box outside the balance to see if any force between them occurs as illustrated in Fig. 4.

#### 2.2.2 Beam Balance (Vertical Measurement)

We developed a very sensitive beam balance using again a frictionless flexural pivot, which features a Cardanic pivot (also using similar flexural pivots) on each side for the counter-weight and the experiment itself as shown in Fig. 5. The beam balance is similar to our torsion balance [40] with a lever arm length of 338 mm. The box uses unannealed mu-metal plates for electrostatic shielding and provides an internal volume of 24.0 x 24.0 x 12.2 cm³. In the case of solenoids, additional three-layer mu-metal shields are wrapped around the coil inside the measurement box.

#### 2.2.3 Torsion Balance (Torque Measurement)

Lastly, a single beam torsion balance is shown in Fig. 6. The measurement box, which consists again of three mu-metal layers for maximum magnetic shielding, is mounted just above the central support which features two frictionless flexural pivots. Due to the maximum load limit of 2 kg on the balance, the measurement volume inside is only 8.0 x 8.0 x 12.0 cm³ for battery, power supply and toroidal coils. The lever arm is 340 mm from the central bearing to mirror from the interferometer.

Actual pictures of the balances and typical configurations for the different measurement boxes are shown in Fig. 7.

### 2.3 Test Articles

Here, we will describe all different test articles. We tried to vary all key aspects and even include some novel configurations to test as many parameters as possible. All dimensions and operational values are summarized in Table 1.

#### 2.3.1 Capacitors

We tested symmetrical disc-shaped capacitor using low, and two different high permittivity dielectrics. Only the highest permittivity capacitor was bought commercially, the others were designed and manufactured by us. Special care was taken to use sufficiently isolated high-voltage connection cables and to coat the finished capacitor with epoxy as shown in Fig. 8. For our $E \times B$ investigation, the capacitor could be mounted between two permanent magnet discs that provided an average field strength of 85 mT as measured with a TLE493D sensor.

Next, we manufactured asymmetrical high-voltage capacitors by using a copper disc for the electrode at one side and the tip of our connection wire with 0.64 mm diameter (AWG22) on the other side as illustrated in Fig. 9. Another variation was to manufacture a symmetrical capacitor with a gradient in the dielectric by using two different (low and high) permittivity inserts in the direction of the electric field [41] as shown in Fig. 10. Still another variation was a capacitor with a trapezoidal-shaped dielectric to create an electric-field gradient perpendicular to the electric field as shown in Fig. 11.

#### 2.3.2 Solenoids

Our basic solenoid was a toroidal coil that traps the magnetic field inside as much as possible. This was important because (static) electric fields can be simply shielded by a Faraday cage, but magnetic fields require multiple high-permeable enclosures that do not provide perfect shielding and significantly increase weight. The toroidal coil was therefore the best design for our tests. Fig. 12 shows our solenoid configuration with twisted wires going to the coil and two layers of wiring (one in clockwise and the other in counter-clockwise direction) that try to achieve a nearly perfect toroidal magnetic field and reduce any single-loop like magnetic field that otherwise would exist. We used 3D printed guides on the top and bottom of our cores for good homogeneity and spacing between the loops.

Similar to the capacitor case, we used a low and a high magnetic permeable core, as well as split-cores made of low and high permeability to create gradients (see Fig. 13). A 50/50 and 75/25 split configuration was chosen in order to test if it makes a difference if the magnetic field crosses the interface only one or twice (with reversed direction). Due to the stray field and possible interaction with Earth's magnetic field, it was not tested on the torsion balance but only on the double-pendulum (which provided more space for shielding) to look for an anomalous force along the gradient direction.

A novel configuration is shown in Fig. 14, where we 3D printed the solenoid core and guiding as one piece out of Polycarbonate (PC). We made a crossed-coil with a classical loop solenoid forming the core of a toroidal coil wrapped around it. This crossed-coil therefore produces two magnetic fields that are perpendicular to each other. We were interested in this configuration as such a non-typical coil has a significant helicity compared to classical coils, where the current helicity is defined as

$$H^J = \int_V \vec{B} \cdot \vec{J} \, dV \tag{5}$$

which is an important parameter for Tokamaks. The magnetic field lines in this case are not closed leading to a chaotic magnetic behavior [42], [43]. It's certainly interesting if helical coils show any difference in our electromagnetic-gravity coupling schemes.

#### 2.3.3 Diodes/Varistors

In order to test tunneling currents, we used Zener diodes and varistors. The Zener diodes were arranged in a large array with all of them pointing in the same direction and operated in reverse direction, where a large tunneling current appears close to its reverse breakdown voltage. Diodes have a thin region where tunneling takes place, therefore we also tested varistors, which consist of a bulk of sintered ZnO grains where tunneling takes place between all grain boundaries providing tunneling through the bulk. Two different low-voltage varistor arrays were built and tested, one with crossing magnetic fields as the capacitors above, as well as a large high-voltage varistor. An example is shown in Fig. 15.

3. Results

All test results for the various configurations are summarized in Table 1, indicating possible vertical or horizontal forces as well as torques as well as the test articles with their measurement orientation. Where applicable, theoretical predictions are given using Ivanov's formula in Equs. (1) and (4). The table also lists our variations for the Mbelek-type configuration using shielded toroidal coils in all different orientation combinations with respect to each other. For all statements, we apply the $3\sigma$ limit rule meaning that every measurement within a $3\sigma$ error bar is noise down to our resolution. In general, our measurements always consider the highest noise level throughout the whole profile and not from a particular sector.

Figs. 16-19 show exemplary measurements of representative capacitor, solenoid and varistor test articles. In particular, the following observations were made within our resolution:

- All types of capacitors tested (low/high-k core dielectric, dielectric gradient, symmetric/asymmetric) did not show any weight change nor did they produce any anomalous force along their electric field or perpendicular to their electric field in case of the trapezoidal capacitor. This is also true for capacitors with a crossed magnetic field where the linear force was assessed along the Poynting vector.
- All toroidal coils tested (low/high core permeability) did not show any weight change nor did they produce any torque along their central symmetry axis. As the volume was quite limited on the torsion balance, tests with the toroidal core using an iron core did show an interaction of their stray field with the environment. Fig. 18 plots the influence of the number of shields on the measured torque and we can see, that each layer is further reducing our measurement. However, after three layers, we were within the weight limit of the balance. We therefore quote our measurement as an upper limit only ($< 2.6 \pm 0.34$ nN-m). If an exponential fit is used, the value for an infinite number of shields would be below our noise level. Still, our upper limit is approximately two orders of magnitude below Ivanov's theory (Equ. 4).
- Toroidal coils with a split-core (low/high permeability) did not produce any linear force along the core's permeability interface direction.
- A helical toroidal crossed-coil did not produce any weight change or torque.
- A pair of toroidal coils, each within a three-layer mu-metal shield, did not produce a force with respect to each other independent of their orientation. Also in this case, iron core toroidal coils slightly interacted with the magnetic shields of the other coils as they had some residual magnetization. We therefore performed the measurements with both coils active and then with each coil active individually to assess their influence on the environment. We then took the value for both on and subtracted the individual influence while leaving the maximum uncertainty as the error bar.
- Tunneling currents produced with Zener diodes or varistors (low and high voltage) did not show any weight change larger than $3\sigma$ nor did they produce any anomalous force along their current direction. This is also true for varistors with a crossed magnetic field where the linear force was assessed along the Poynting vector. These measurements were in particular challenging as the varistors heated up quite substantially creating a lot of outgassing that was interfering with the measurements and required long cooling down periods between each measurement. An example is shown in Fig. 19 for both beam and double-pendulum balance measurements. The double-pendulum seems to show an effect of $-25 \pm 5.3$ nN, which would be outside of our $3\sigma$ noise limit. We tried to limit the outgassing effect as much as possible by providing a rather thick epoxy layer on the varistors, however, the amount of power that was going through them was so high that outgassing still happened. As the power was more than 13 Watts, the equivalent photon pressure force is 44 nN, which is well above our measurement value. We therefore consider also this test within at least photon pressure noise, which compares an anomalous force to present state-of-the-art propellantless propulsion. Future tests may improve on this aspect. Varistors with higher voltage and less current (and therefore less power), did not show an anomaly, also on the double-pendulum balance.

## 4. Conclusion

No anomalous forces or torques were found in our in-depth search for a coupling between electromagnetism and gravity using steady fields. Extending our previous results, we can now rule out Ivanov's theory by at least two- to three orders of magnitude, even for the case of low permittivity dielectrics and solenoids. Several configurations were tested for the first time like toroidal and cross-coils and tunneling currents. Our data provides strong limits for theories, claimed anomalous forces and can serve as the basis for future experiments. As we only studied static fields, the next step is to explore couplings in the high-frequency domain.

| | Maxwell | | | | | | Non-Maxwell |
|---|---|---|---|---|---|---|---|
| | Electric Field **E** | Magnetic Field **B** | **ExB**, EM-Wave | Charge **q** | Current **I** | Speed of Light **c** Vacuum $\varepsilon_0,\mu_0$ | Additional Scalar Field Magnetic Monopoles Born-Infeld EM, … |
| Gravitational Field **g** | Ivanov Root Gravity* | Ivanov Root Gravity | EMDrive* Feigel Effect and Vacuum MHD | | Superconducting*, Tunneling or Divergent Currents (Biefeld-Brown)* | Vacuum/Casimir Forces | Mbelek-Scalar Force* |
| Mass **m** | Assis-Weber Gravity Model and Electric Polarization* | Alzofon Nuclear Spin Polarization* | | Inertial Mass from Stationary Charges* Saxl-Type, Kaluza-Klein* | | ZPE Mass Models Curie-type Critical Temperature for Spacetime* | |

* Experimental results available in literature

Fig. 1 Coupling-Schemes of Gravity and Electromagnetism including Theory and Experimental Claims (Crossed Cell marks Incompatibility between Vector and Scalar Quantities, Green-Areas are Assessed in Present Work)

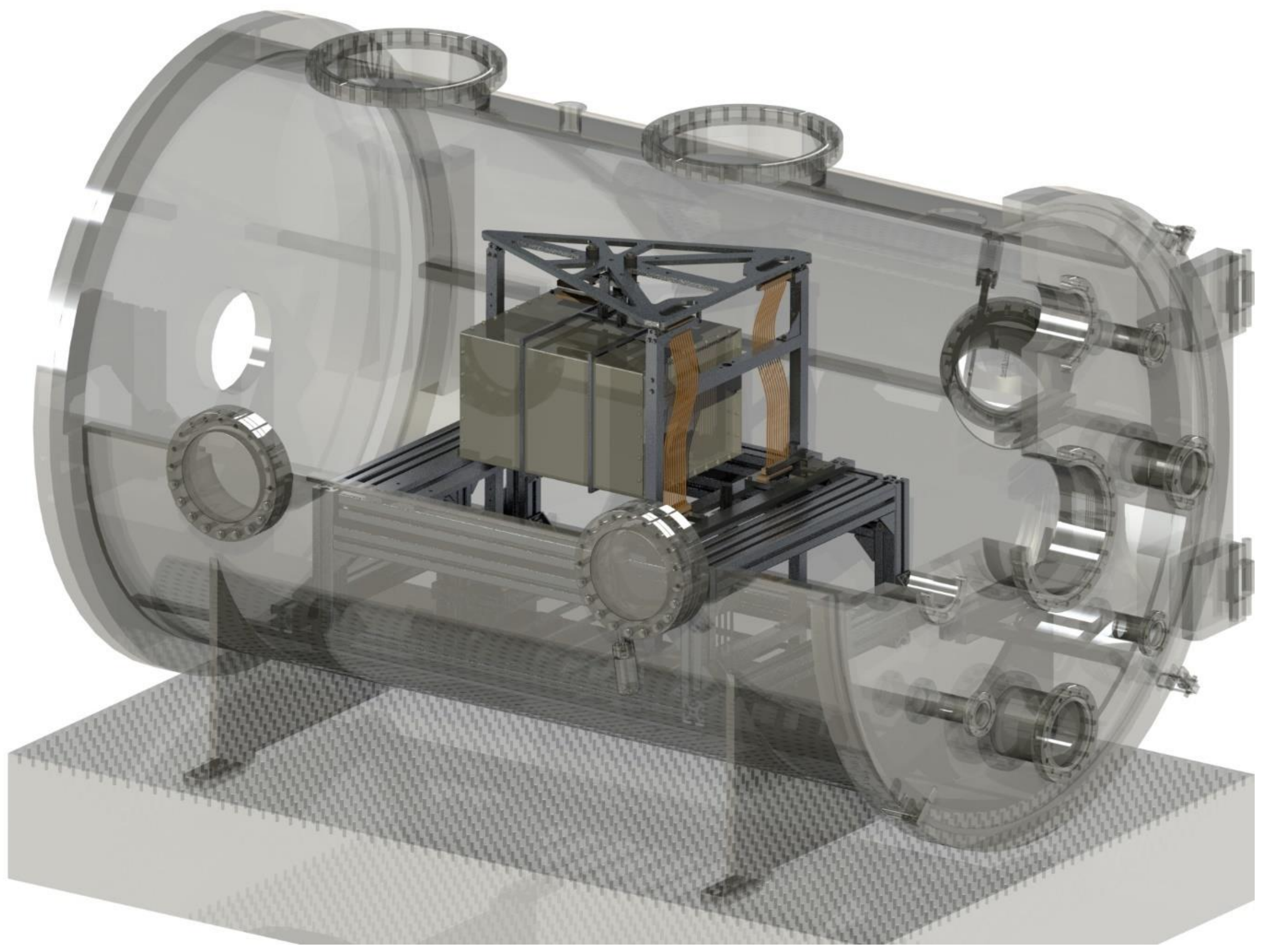

Fig. 2 Overall Setup – Double Pendulum Balance with Mu-Metal Box inside of Vacuum Chamber

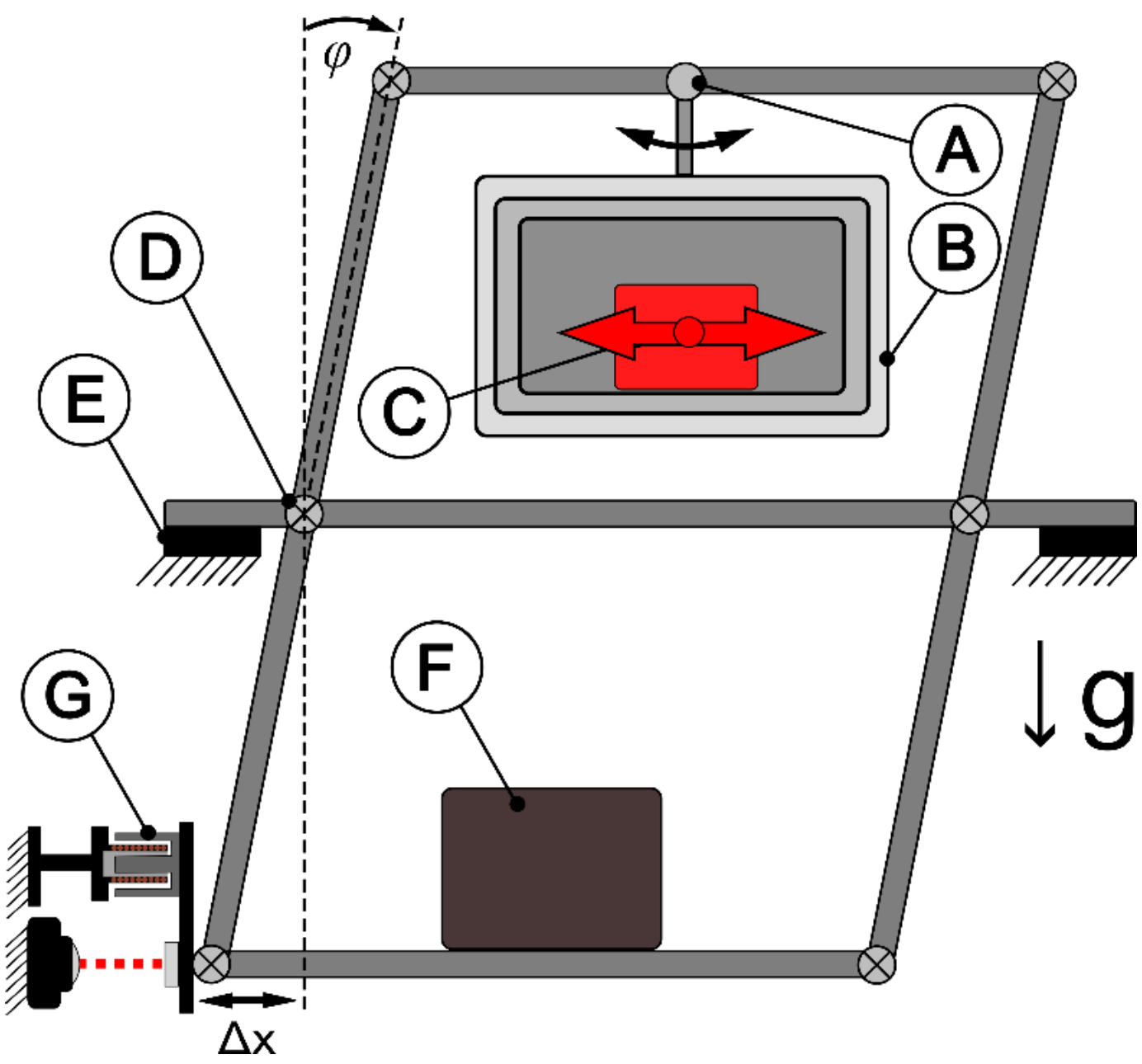

Fig. 3 Double-Pendulum Setup (A..Pivot, B..Mu-Metal Shields, C..Test Article, D..Frictionless Bearing, E...Support, F..Counter-Weight, G..Voice-Coil Calibrator and Interferometer Mirror)

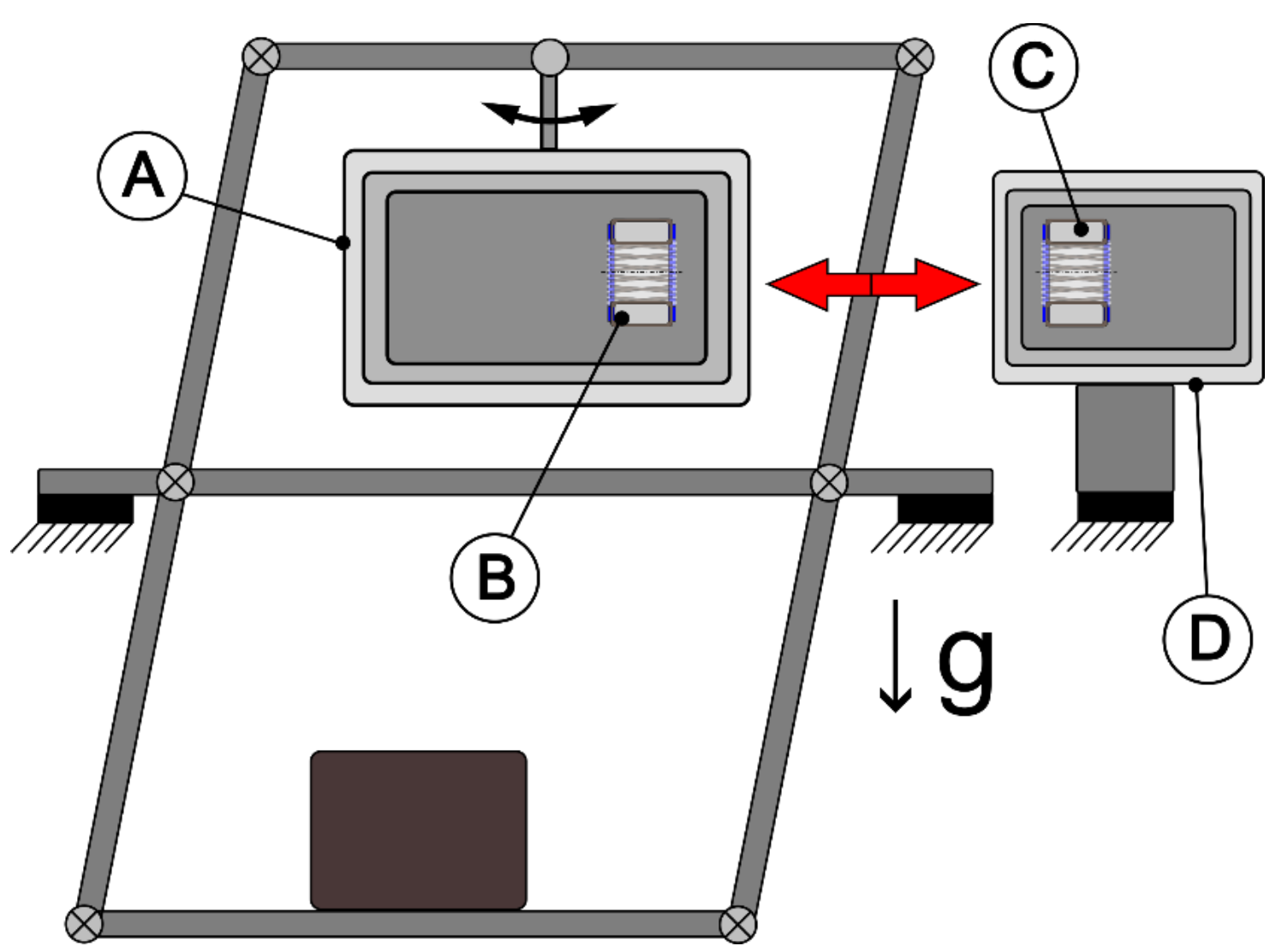

Fig. 4 Mbelek-Type Coil-Coil Setup using the Double-Pendulum to Investigate Force between Two Toroidal Coils (both coils in horizontal orientation) (A.. Mu-Metal Shields on Balance, B..Coil on Balance, C..External Coil, D..External Mu-Metal Shields)

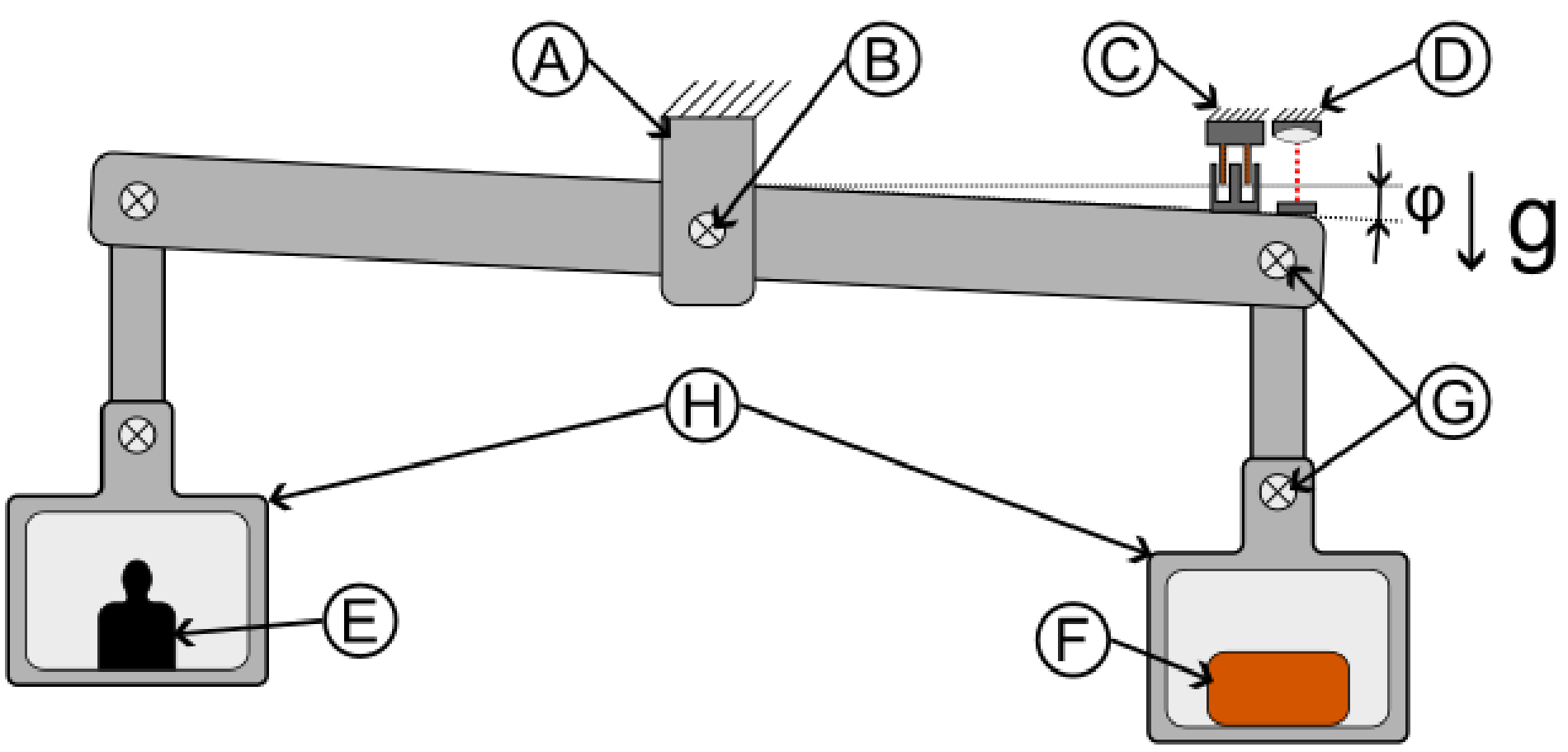

Fig. 5 Beam-Balance Setup (A..Support, B..Frictionless Bearing, C,D..Voice-Coil Calibrator and Interferometer Mirror, E..Counter-Weight, F..Test Article, G..Cardanic Pivot, H..Unannealed Mu-Metal)

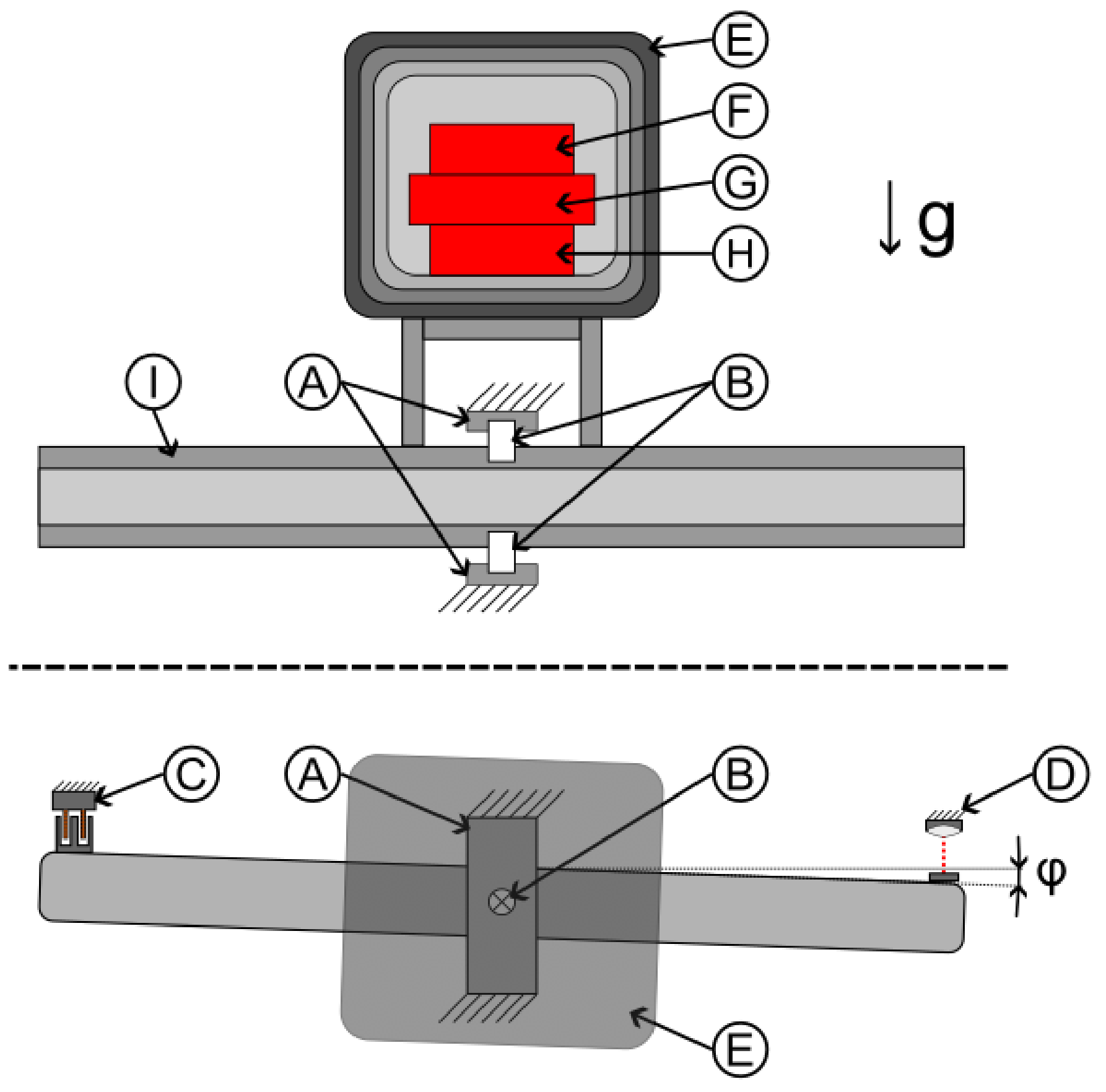

Fig. 6 Torsion-Balance Setup: Top-Side View, Bottom-Top View (A..Support, B..Frictionless Bearing, C..Voice-Coil Calibrator, D..Interferometer Mirror, E..Measurement Box, F..Battery, G..Power Supply, H..Solenoid, I..Beam)

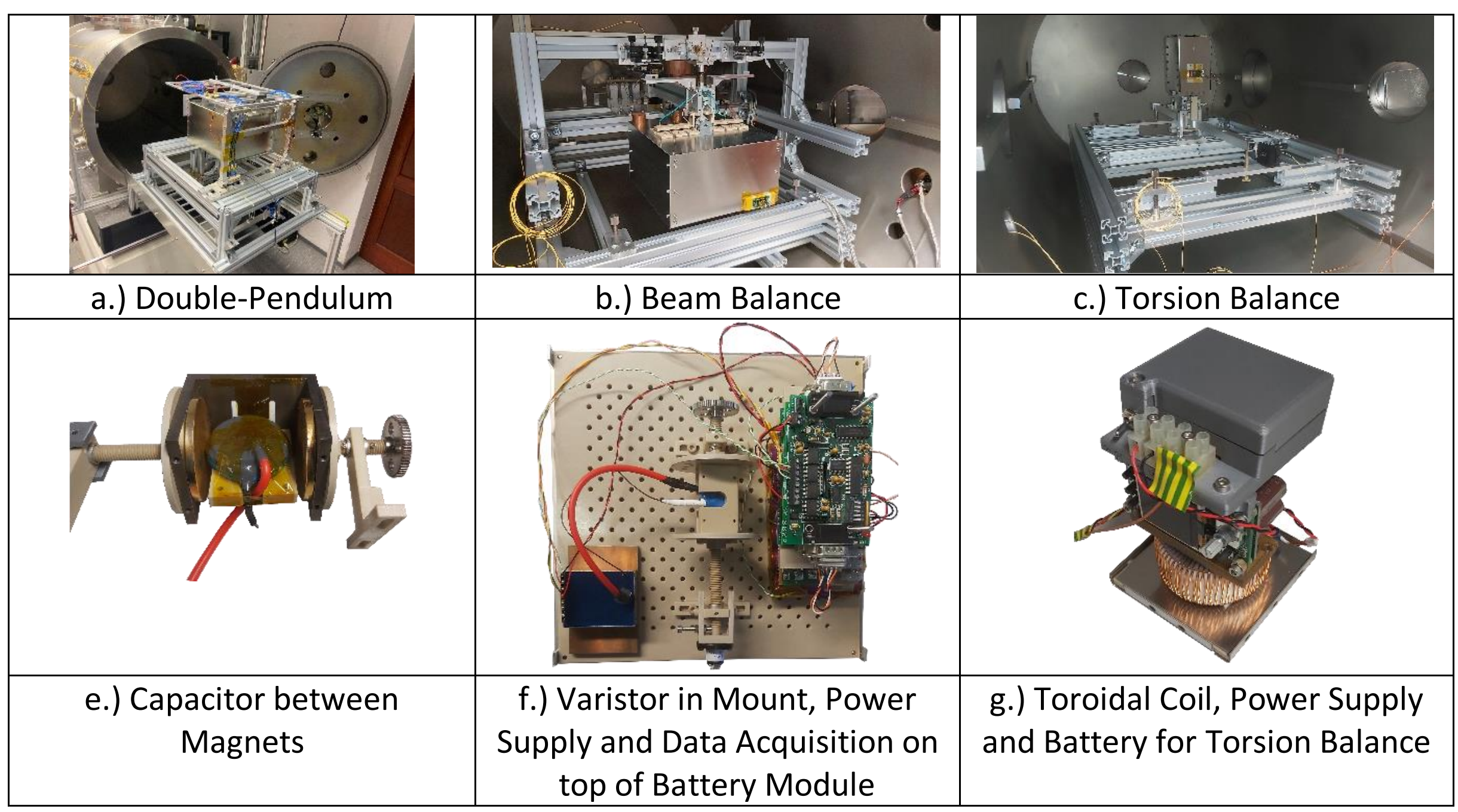

a.) Double-Pendulum | b.) Beam Balance | c.) Torsion Balance

e.) Capacitor between Magnets | f.) Varistor in Mount, Power Supply and Data Acquisition on top of Battery Module | g.) Toroidal Coil, Power Supply and Battery for Torsion Balance

Fig. 7 Typical Configuration of Balance and Measurement Box Details

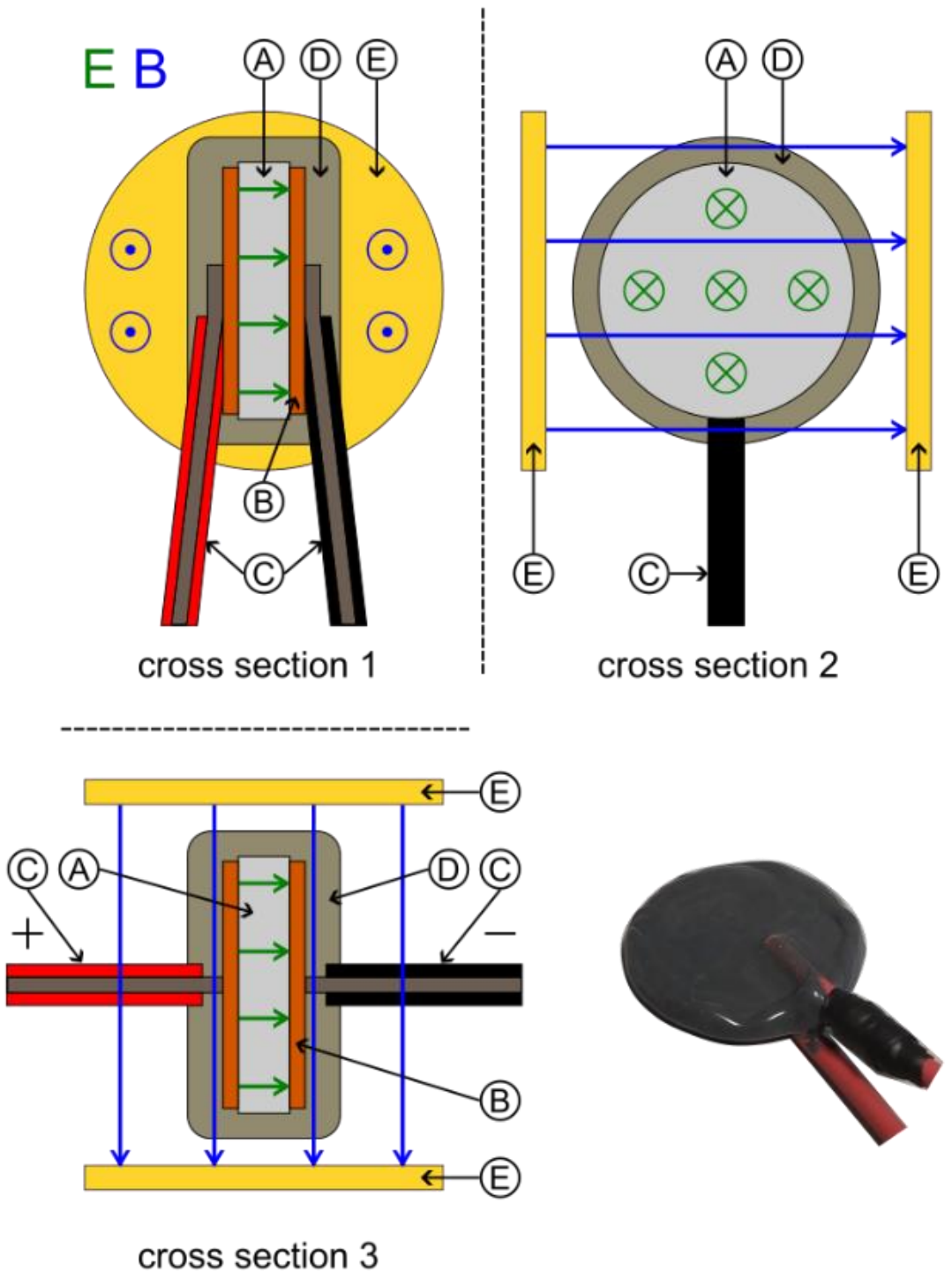

Fig. 8 Schematic Configuration of Symmetrical High-Voltage Capacitor (A..Dielectric, B..Conductors, C..Connection Wires, D..Isolation Epoxy, E..Optional Permanent Magnets)

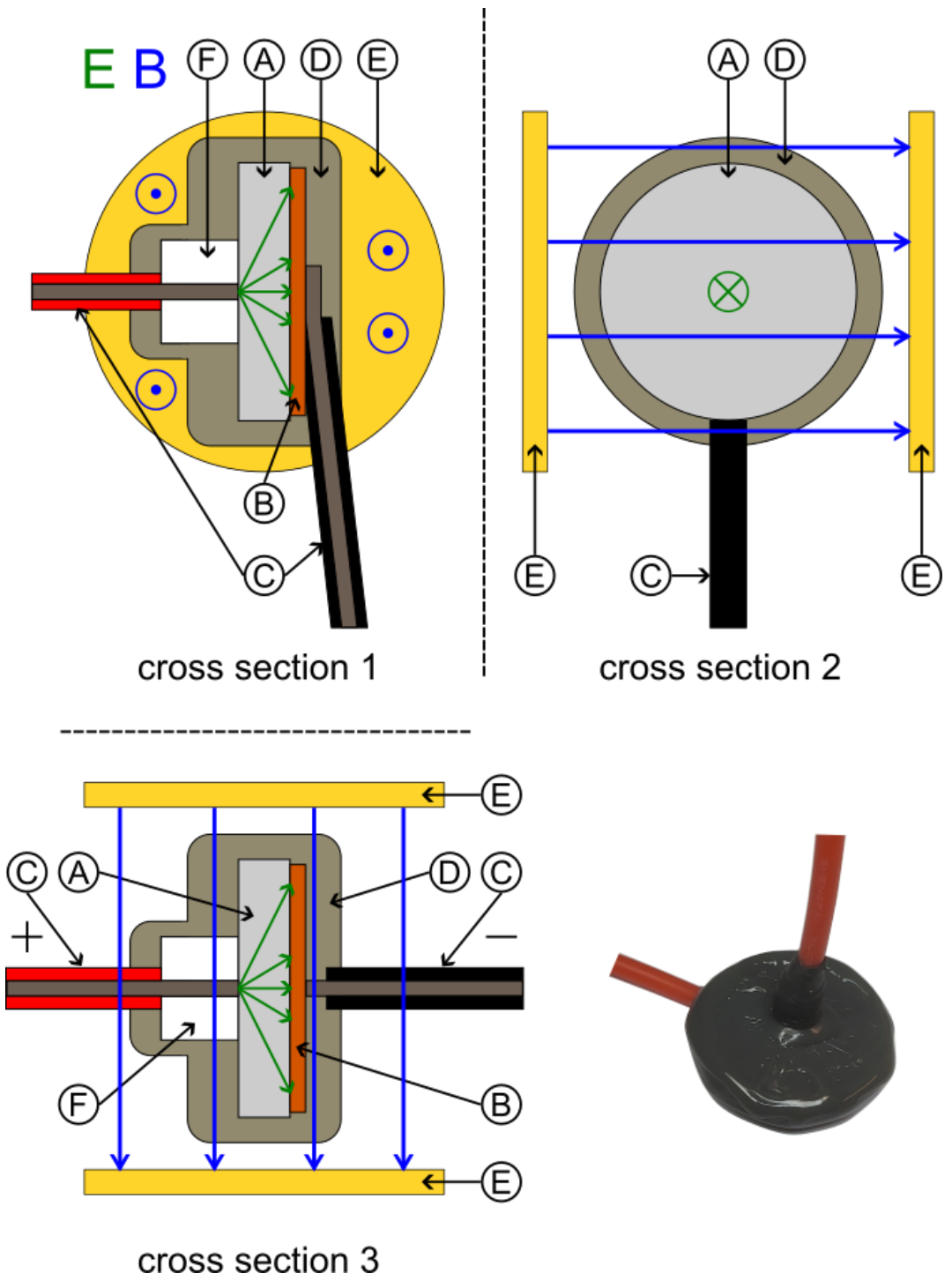

Fig. 9 Schematic Configuration of Asymmetrical High-Voltage Capacitor (A..Dielectric, B..Conductors, C..Connection Wires, D..Isolation Epoxy, E..Optional Permanent Magnets, F..Isolator)

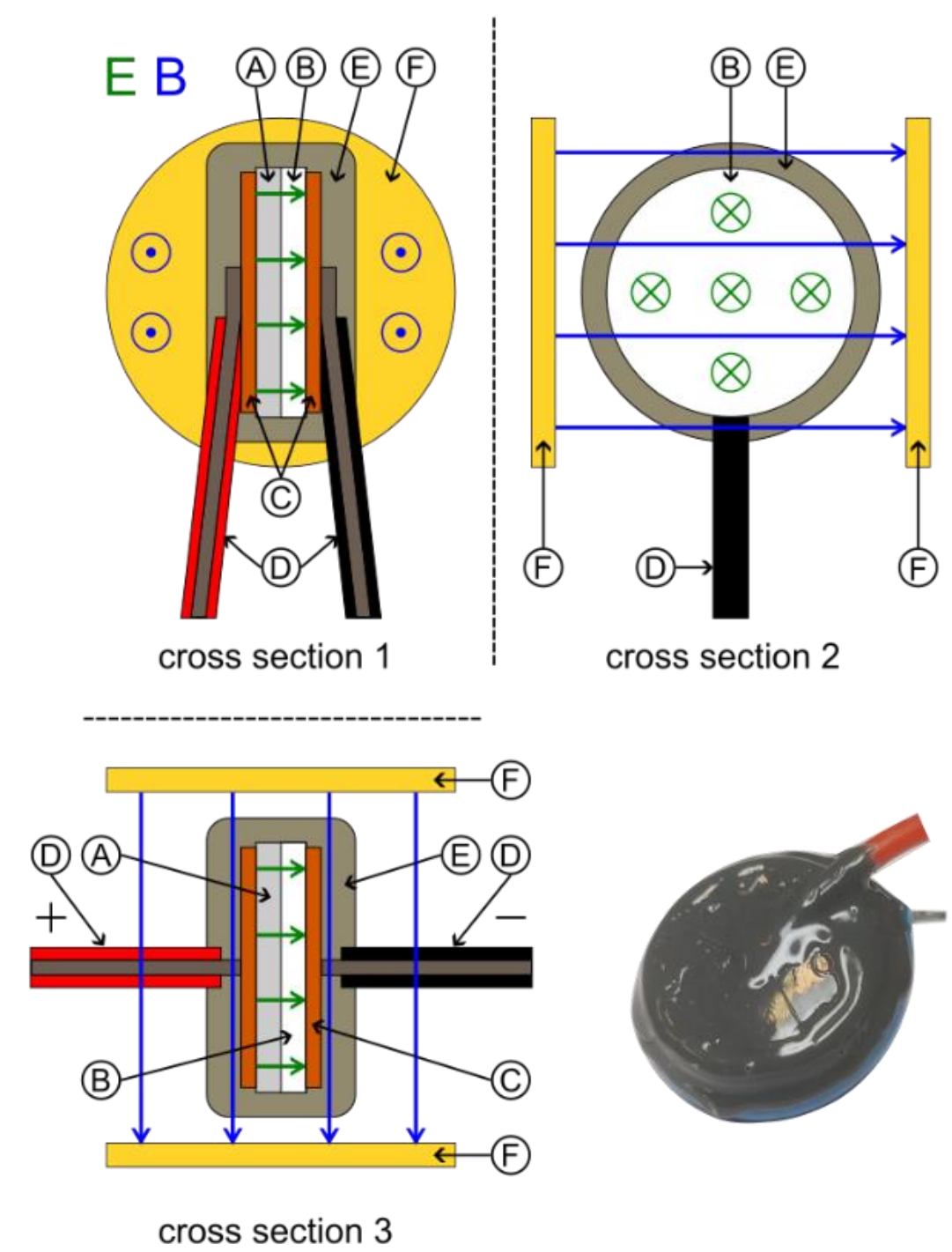

Fig. 10 Schematic Configuration of Gradient High-Voltage Capacitor (A..Dielectric 1, B..Dielectric 2, C..Conductors, D..Connection Wires, E..Isolation Epoxy, F..Optional Permanent Magnets)

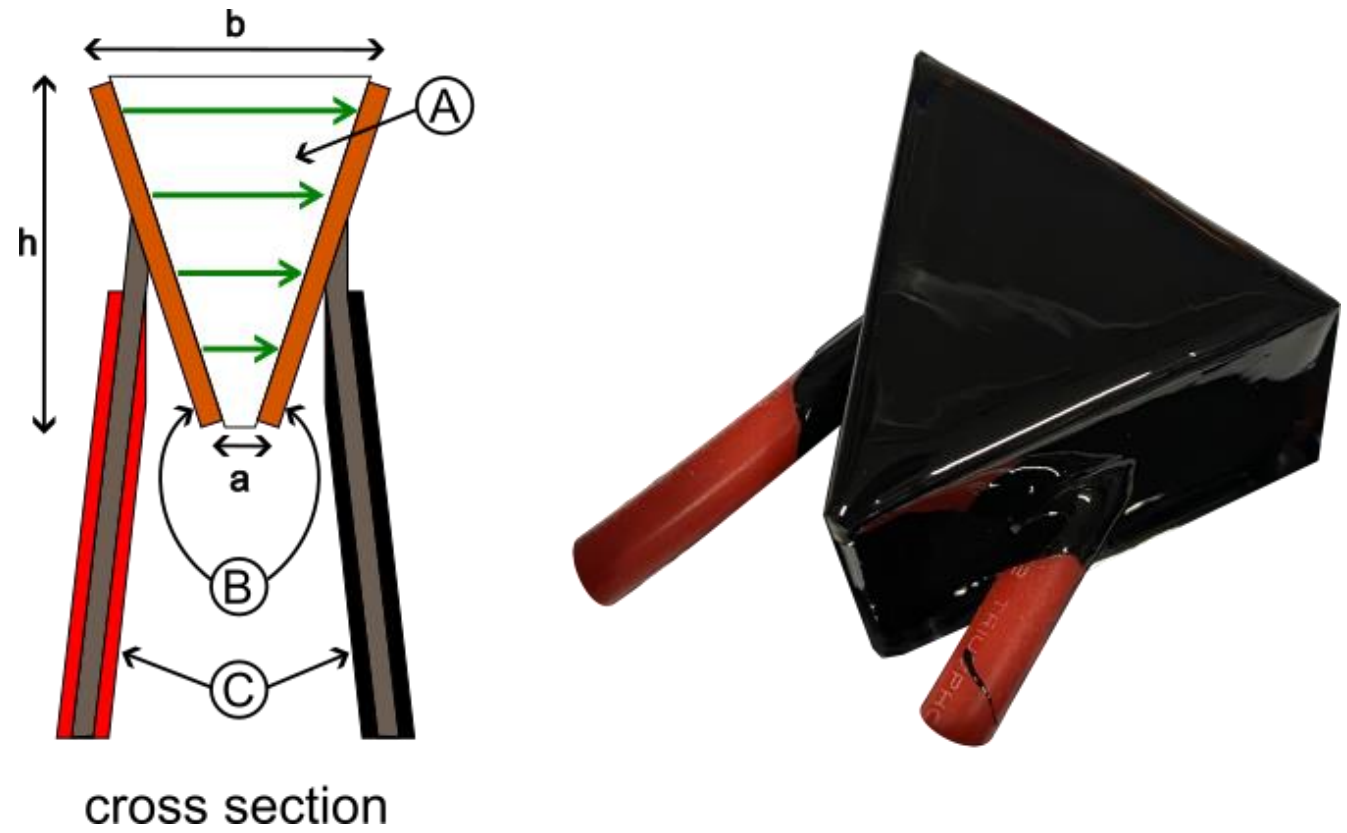

Fig. 11 Schematic Configuration of Trapezoidal-Shaped Asymmetric High-Voltage Capacitor (A..Dielectric, B..Conductors, C..Connection Wires, a..Short Base, b..Long Base, h..Height)

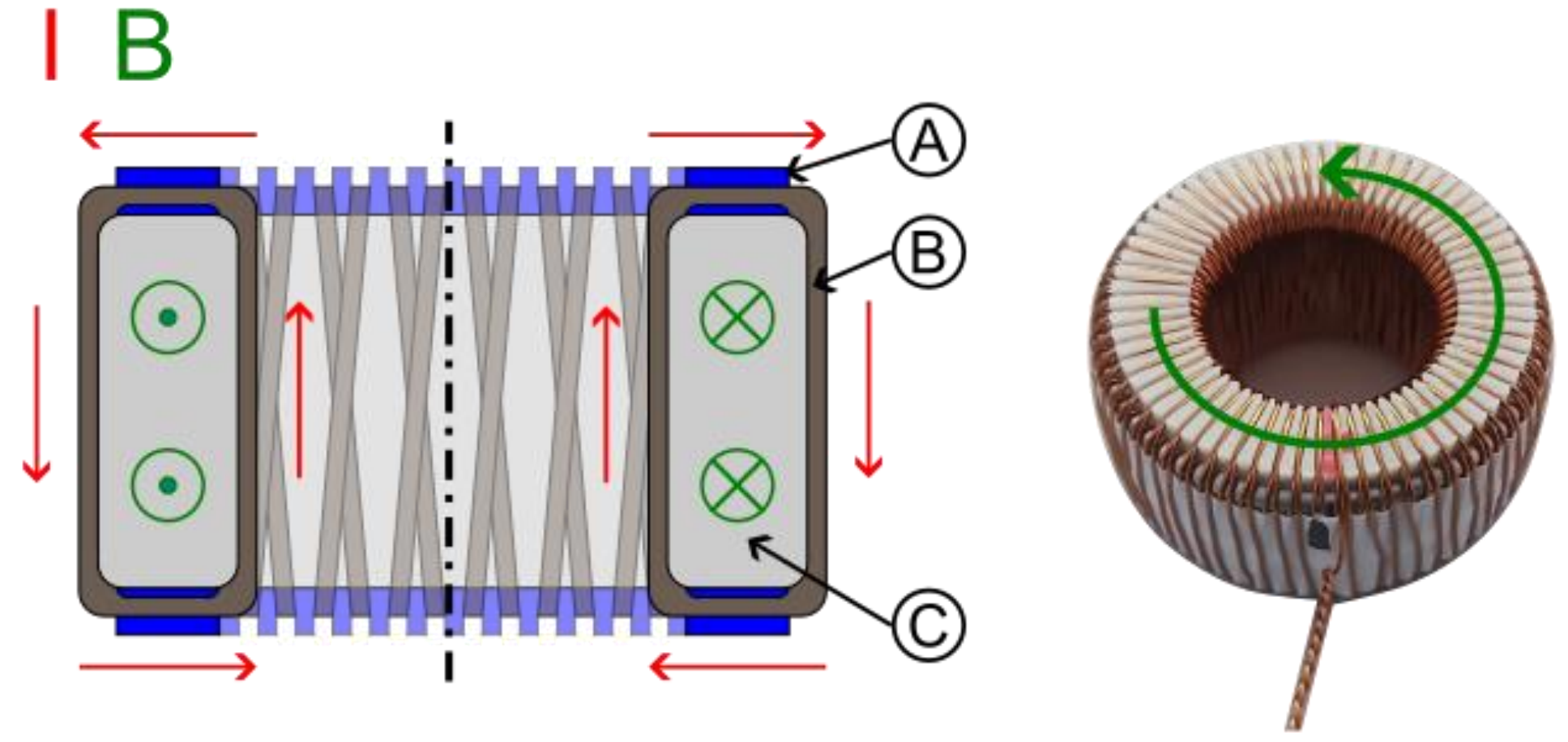

Fig. 12 Schematic Configuration of Toroidal Coil (A..3D-Printed Guide, B..Wire, C..Core)

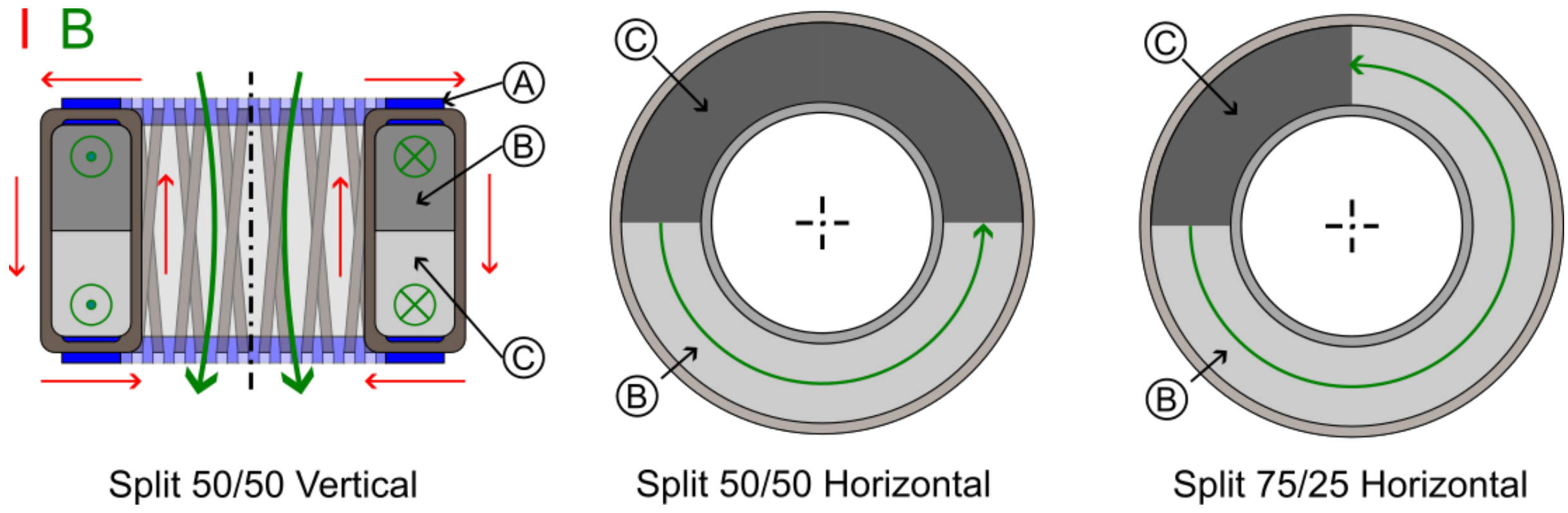

Fig. 13 Schematic Configuration of Split-Core Toroidal Coils (A..3D-Printed Guide, B..Core 1, C..Core 2)

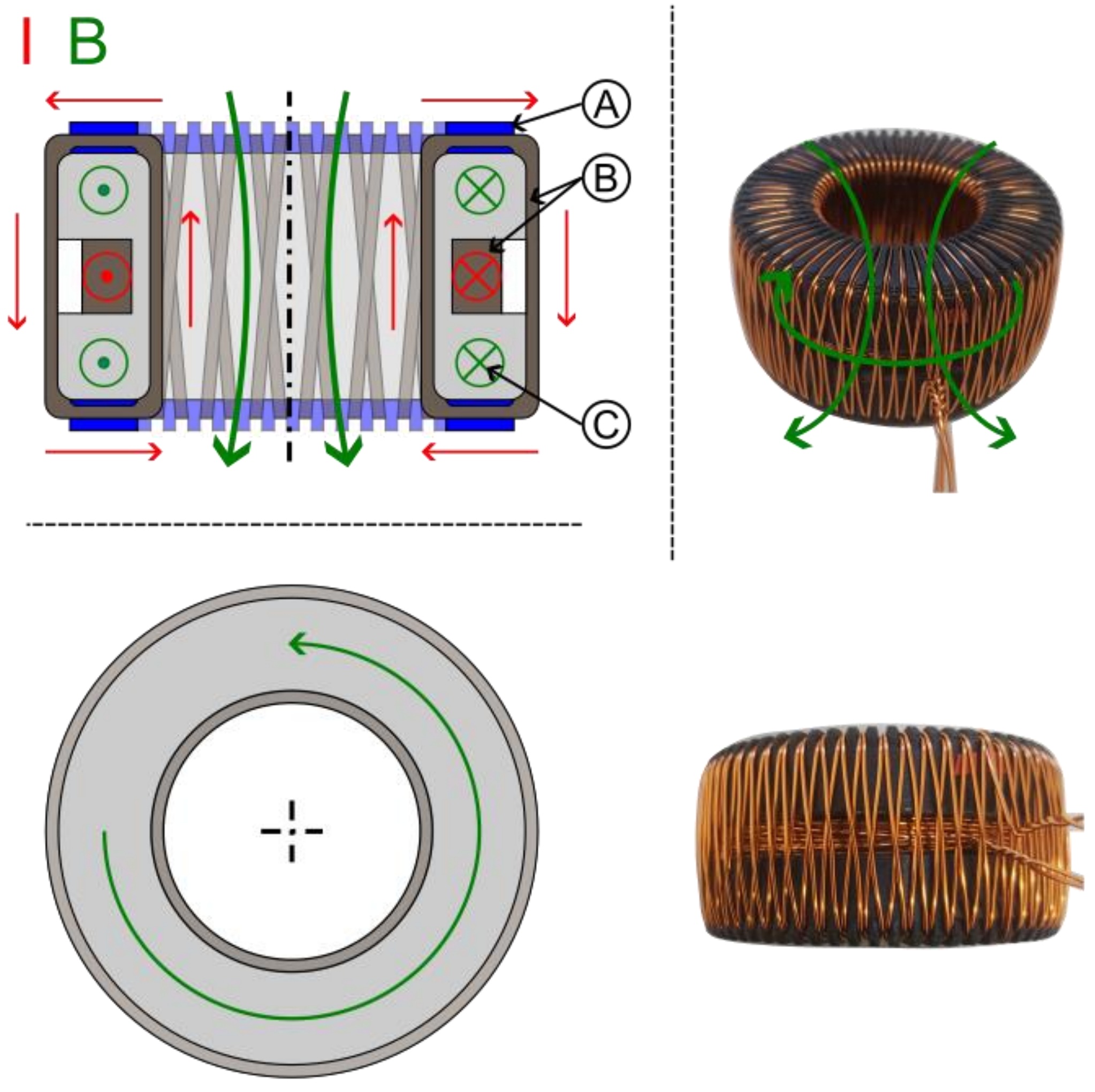

Fig. 14 Schematic Configuration of Crossed Coil (A..3D-Printed Guide, B..Wires, C..Core)

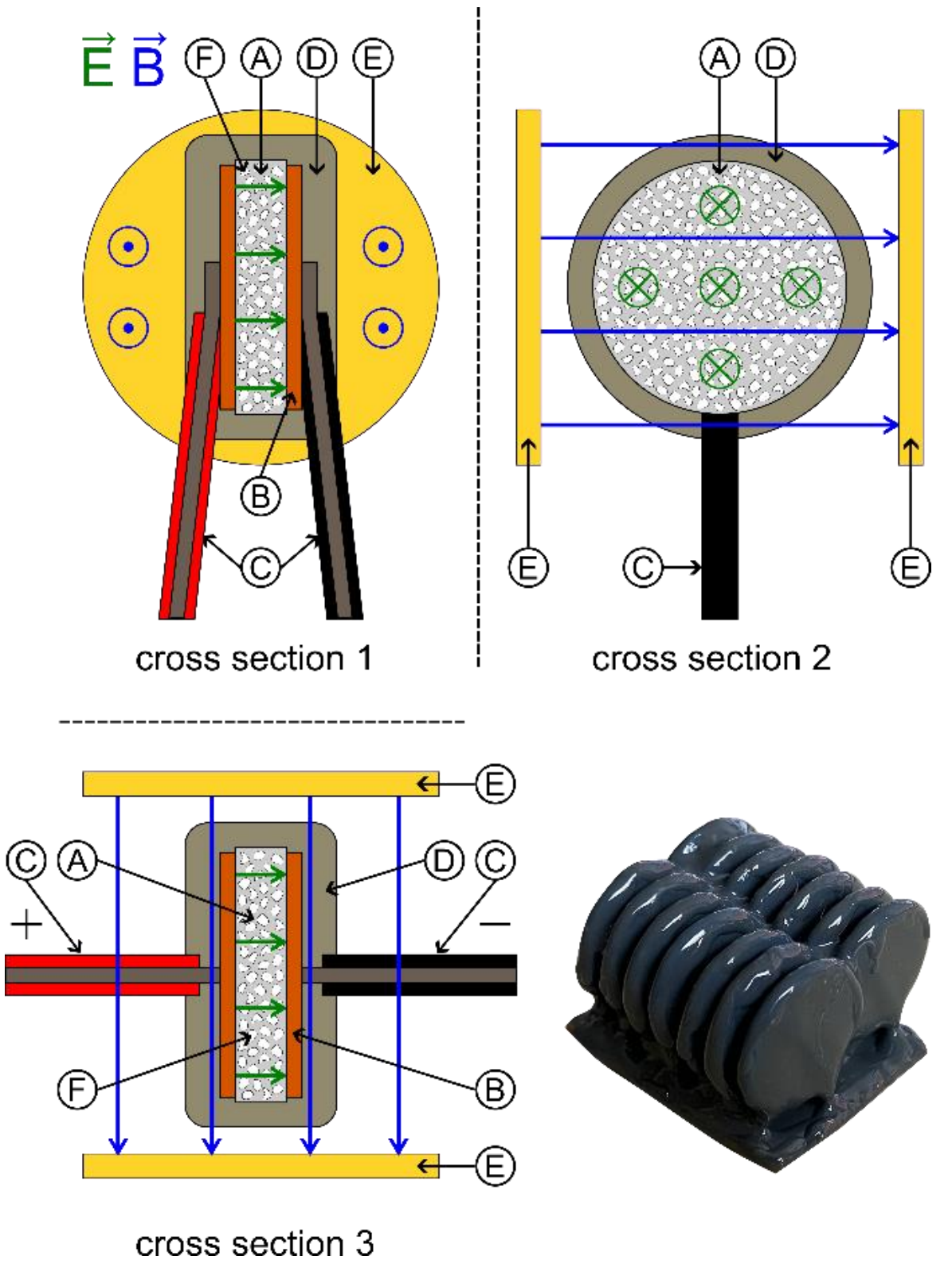

Fig. 15 Schematic Configuration of Single Varistor (A..ZnO Grains, B..Conductors, C..Connection Wires, D..Isolation Epoxy, E..Optional Permanent Magnets) and 14xVaristor Stack Example

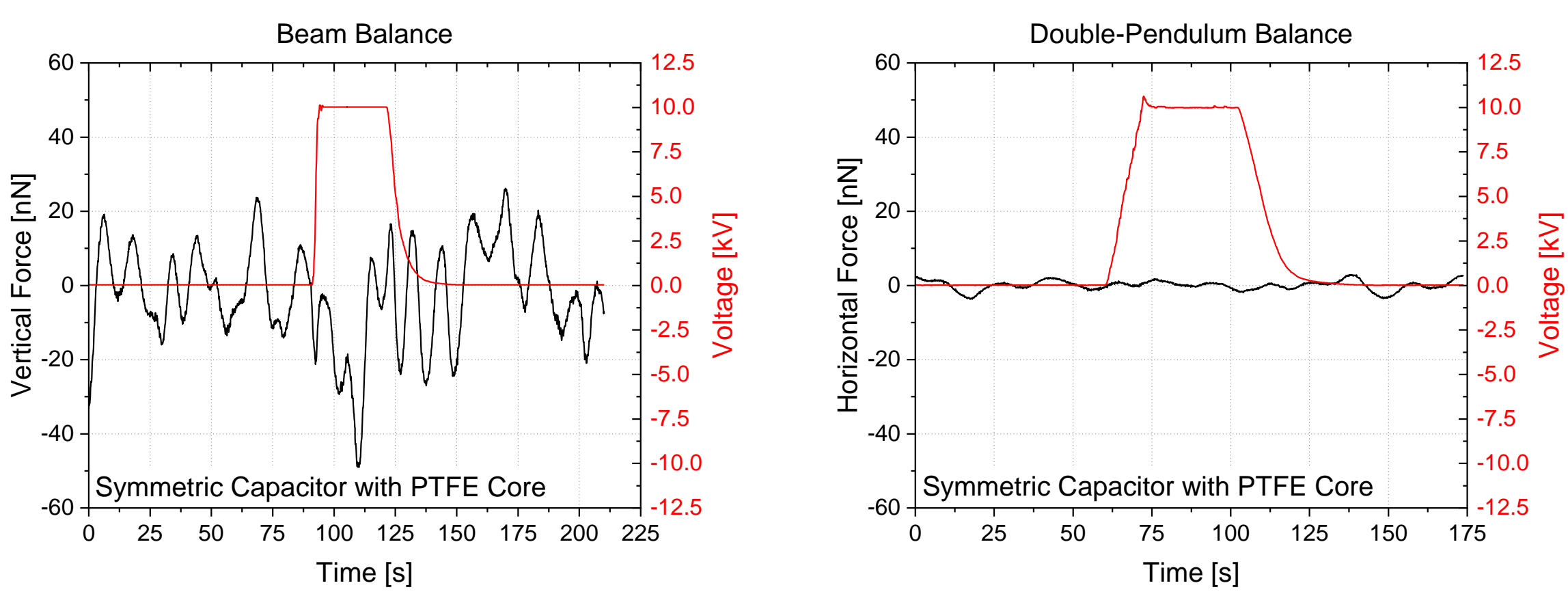

Fig. 16 Symmetric PTFE Capacitor Measurement with Beam Balance (left) and Double-Pendulum Balance (right)

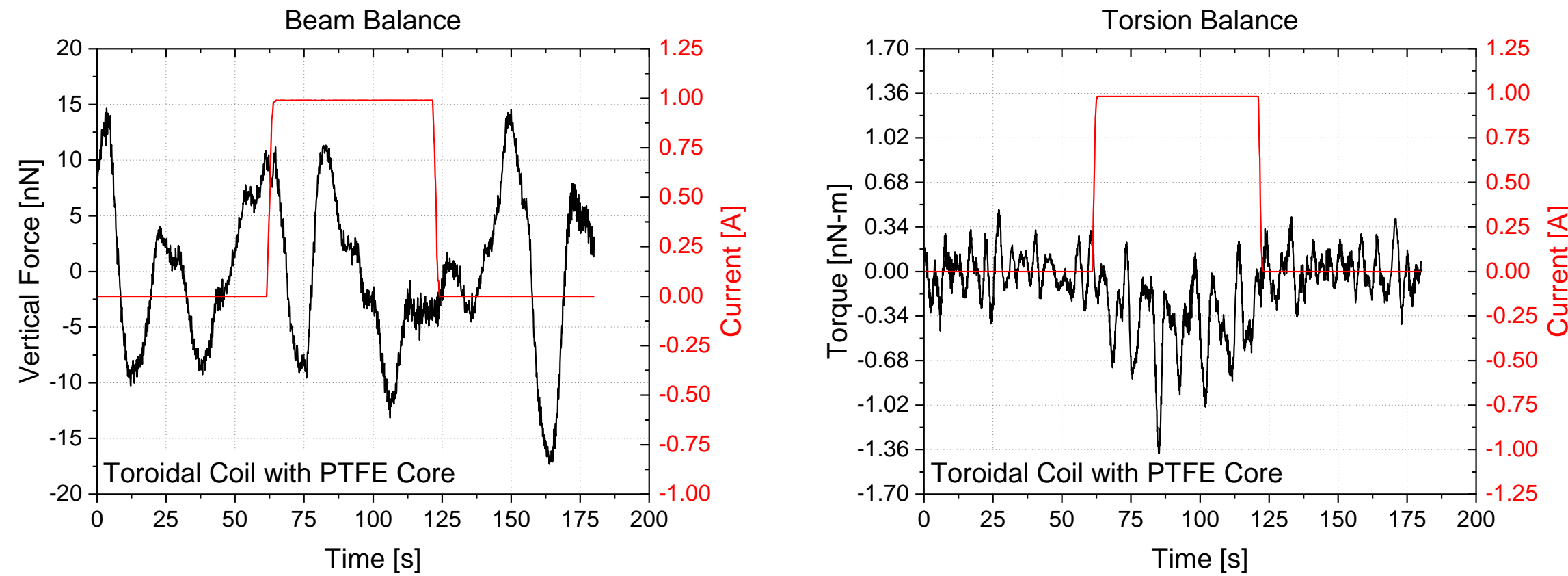

Fig. 17 Toroidal Coil with PTFE Core Measurement with Beam Balance (left) and Torsion Balance (right)

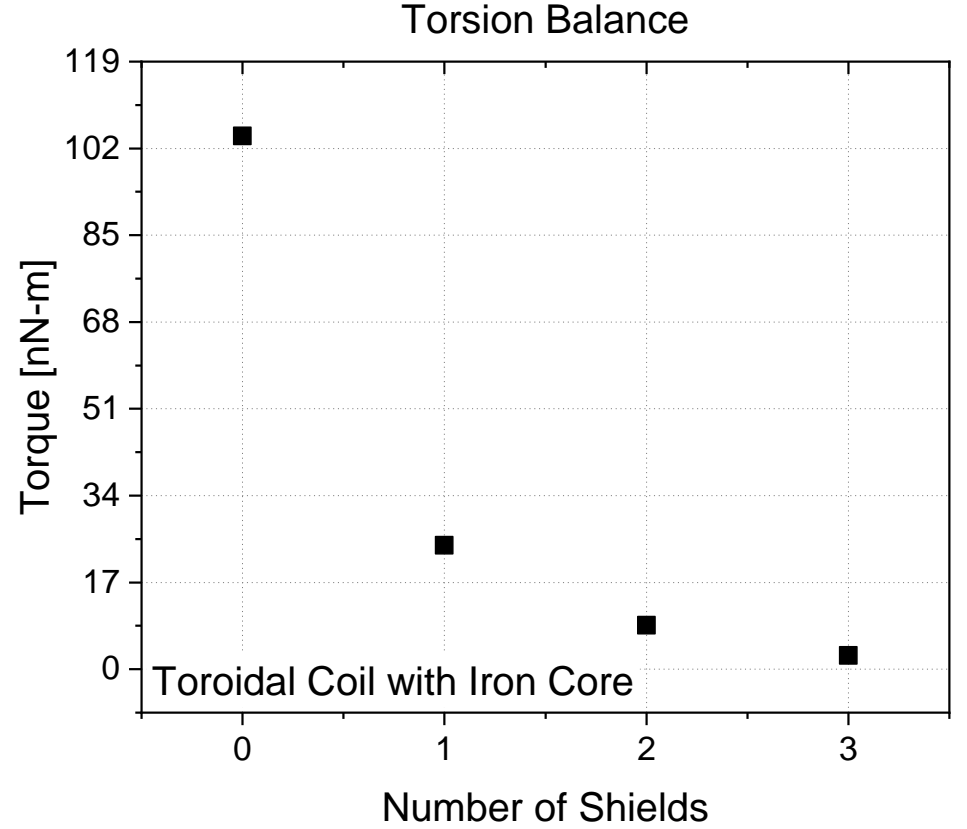

Fig. 18 Influence of Number of Mu-Metal Shields on Torque Measurement with Torsion Balance and Toroidal Coil with Iron Core

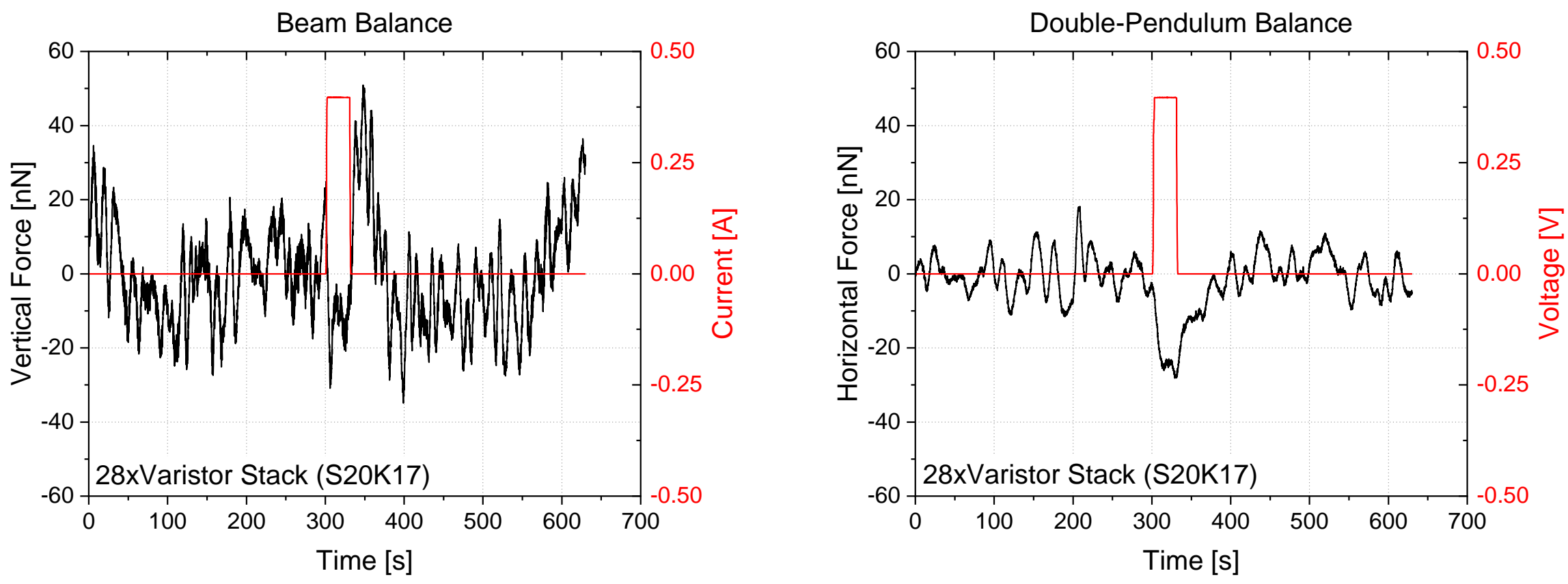

Fig. 19 Varistor Stack Measurement with Beam Balance (left) and Double-Pendulum Balance (right)

| Test Article | Type (Commercial Name if available) | Core | Dimensions | Electric Parameter | Measurement Orientation | Ivanov Theory [nN] | Vertical Force [nN] | Horizontal Force [nN] | Torque [nN-m] |
|---|---|---|---|---|---|---|---|---|---|
| Capacitor | | | | | | | | | |
| | Symmetric | PTFE ($\varepsilon_r$ = 2.1), $\rho$ = 2.2 g/cm$^3$ | d = 35 mm, t = 1.5 mm | 10.0 kV | E | 746 nN | -17.9 ± 18.8 | 0.39 ± 0.66 | N/A |
| | | PZT-5H* ($\varepsilon_r$ = 3500), $\rho$ = 7.8 g/cm$^3$ | d = 35 mm, t = 4.0 mm | 7.0 kV | E | 75,549 nN | 9.6 ± 7.1 | 0.22 ± 0.49 | N/A |
| | (HVC-50kV-D40-F20-332M) | Y5T** ($\varepsilon_r$ = 6000), $\rho$ = 3.9 g/cm$^3$ | d = 40 mm, t = 20.0 mm | 30 kV | E | 423,931 nN | -9.5 ± 13.5 | N/A | N/A |
| | (HVC-50kV-D40-F20-332M) | Y5T** ($\varepsilon_r$ = 6000), $\rho$ = 3.9 g/cm$^3$ | d = 40 mm, t = 20.0 mm | 20 kV | E | 282,620 nN | N/A | 0.6 ± 2.2 | N/A |
| | Symmetric $\perp$ B-Field (85 mT) | PTFE ($\varepsilon_r$ = 2.1), $\rho$ = 2.2 g/cm$^3$ | d = 35 mm, t = 1.5 mm | 10.0 kV | E×B | | -7.1 ± 14.3 | N/A | N/A |
| | | PZT-5H* ($\varepsilon_r$ = 3500), $\rho$ = 7.8 g/cm$^3$ | d = 35 mm, t = 4.0 mm | 7.0 kV | E×B | | 18.0 ± 28.0 | N/A | N/A |
| | (HVC-50kV-D40-F20-332M) | Y5T** ($\varepsilon_r$ = 6000), $\rho$ = 3.9 g/cm$^3$ | d = 40 mm, t = 19.0 mm | 10.0 kV | E×B | | 16.9 ± 18.3 | N/A | N/A |
| | Asymmetric | PTFE ($\varepsilon_r$ = 2.1), $\rho$ = 2.2 g/cm$^3$ | d = 35 mm, t = 1.5 mm | 10.0 kV | E | | 6.8 ± 7.7 | 4.3 ± 3.4 | N/A |
| | | PZT-5H* ($\varepsilon_r$ = 3500), $\rho$ = 7.8 g/cm$^3$ | d = 35 mm, t = 4.0 mm | 2.0 kV | E | | -1.2 ± 16.7 | 0.7 ± 1.7 | N/A |
| | Asymmetric $\perp$ B-Field (85 mT) | PTFE ($\varepsilon_r$ = 2.1), $\rho$ = 2.2 g/cm$^3$ | d = 35 mm, t = 1.5 mm | 10.0 kV | E×B | | -11.0 ± 8.3 | N/A | N/A |
| | | PZT-5H* ($\varepsilon_r$ = 3500), $\rho$ = 7.8 g/cm$^3$ | d = 35 mm, t = 4.0 mm | 2.0 kV | E×B | | -20.6 ± 13.2 | N/A | N/A |
| | Gradient Dielectric 50/50 | PTFE ($\varepsilon_r$ = 2.1), PZT-5H* ($\varepsilon_r$ = 3500) | d = 35 mm, t = (4.0 + 4.0) mm | 7.5 kV | E (=Gradient $\varepsilon_r$) | | 2.7 ± 4.6 | 0.2 ± 0.6 | N/A |
| | Grad. Diel. 50/50 $\perp$ B-Field (85 mT) | PTFE ($\varepsilon_r$ = 2.1), PZT-5H* ($\varepsilon_r$ = 3500) | d = 35 mm, t = (4.0 + 4.0) mm | 7.5 kV | E×B | | -11.0 ± 14.7 | N/A | N/A |
| | Asym. Dielectric (Trapezoid) | PTFE ($\varepsilon_r$ = 2.1) | h = 40 mm, a = 2 mm, b = 40 mm | 20 kV | $\perp$ to E | | N/A | 1.6 ± 4.3 | N/A |
| Coil | | | | | | | | | |
| | Toroidal | PTFE ($\mu_r$ = 1), r = 2.2 g/cm$^3$ | $d_o/d_i$ = 58/30 mm, h = 29 mm, N = 130 | 1 A | C-axis, rot(B) | 2.4 nN-m | -4.1 ± 4.1 | N/A | -0.37 ± 0.27 |
| | | Iron ($\mu_r$ = 500), $\rho$ = 7.8 g/cm$^3$ | $d_o/d_i$ = 58/30 mm, h = 29 mm, N = 130 | 1 A | rot(B) | 187.8 nN-m | 7.0 ± 6.4 | N/A | < 2.6 ± 0.34 |
| | Gradient 50/50 Vertical | PTFE ($\mu_r$ = 1), Iron ($\mu_r$ = 500) | $d_o/d_i$ = 58/30 mm, h = 29 mm, N = 130 | 1 A | Gradient $\mu_r$ | | N/A | 1.1 ± 1.1 | N/A |
| | Gradient 50/50 Horizontal | PTFE ($\mu_r$ = 1), Iron ($\mu_r$ = 500) | $d_o/d_i$ = 58/30 mm, h = 29 mm, N = 130 | 1 A | Gradient $\mu_r$ | | N/A | 2.1 ± 1.8 | N/A |
| | Gradient 75/25 Vertical | PTFE ($\mu_r$ = 1), Iron ($\mu_r$ = 500) | $d_o/d_i$ = 58/30 mm, h = 29 mm, N = 130 | 1 A | Gradient $\mu_r$ | | N/A | 1.3 ± 3.2 | N/A |
| | Crossed-Coil | PC ($\mu_r$ = 1), $\rho$ = 1.2 g/cm$^3$ | $d_o/d_i/d_c$ = 58/30/41.5 mm, h = 29 mm,<br>$N_t$=130, $N_c$=75 | 1 A | C-axis, rot($B_{coil}$) | | 13.8 ± 6.8 | N/A | 0.34 ± 0.92 |
| Toroidal Coil-Toroidal Coil (Distance from Central Point-Central Point) | | | | | | | | | |
| | Distance = 138 mm | PLA (Central Axes = V/V) | $d_o/d_i$ = 58/30 mm, h = 29 mm, N = 130 | 2 A | rot(B) vs. rot(B) | | N/A | -0.2 ± 2.7 | N/A |
| | Distance = 123.5 mm | PLA (Central Axes = H/V) | $d_o/d_i$ = 58/30 mm, h = 29 mm, N = 130 | 2 A | rot(B) vs. C-axis | | N/A | -1.9 ± 3.4 | N/A |
| | Distance = 109 mm | PLA (Central Axes = H/H) | $d_o/d_i$ = 58/30 mm, h = 29 mm, N = 130 | 2A | C-axis vs. C-axis | | N/A | -0.6 ± 2.8 | N/A |
| | Distance = 138 mm | Fe (Central Axes = V/V) | $d_o/d_i$ = 58/30 mm, h = 29 mm, N = 130 | 2 A | rot(B) vs. rot(B) | | N/A | 2 ± 12 | N/A |
| | Distance = 123.5 mm | Fe (Central Axes = H/V) | $d_o/d_i$ = 58/30 mm, h = 29 mm, N = 130 | 2 A | rot(B) vs. C-axis | | N/A | 3 ± 10 | N/A |
| | Distance = 109 mm | Fe (Central Axes = H/H) | $d_o/d_i$ = 58/30 mm, h = 29 mm, N = 130 | 2 A | C-axis vs. C-axis | | N/A | 4 ± 11 | N/A |
| Varistor | | | | | | | | | |
| | Parallel (S20K17) | ZnO | d = 21.5 mm, h = 4.8 mm, N = 28 | 33 V, 0.4 A | I | | -12.3 ± 11 | -25 ± 5.3*** | N/A |
| | Parallel (S20K17) $\perp$ B-Field (85 mT) | ZnO | d = 21.5 mm, h = 4.8 mm, N = 14 | 33 V, 0.3 A | I×B | | 17.1 ± 8.7 | N/A | N/A |
| | Single (S20K1000) | ZnO | d = 21.5 mm, h = 11.4 mm, N = 1 | 1850 V, 1 mA | I | | -0.8 ± 5.5 | N/A | N/A |
| | Single $\perp$ B-Field (85 mT) (S20K1000) | ZnO | d = 21.5 mm, h = 11.4 mm, N =1 | 1740 V, 0.83 mA | I×B | | 1.8 ± 15 | N/A | N/A |
| | Single (V282BB60) | ZnO | d = 60.0 mm, h = 28 mm, N = 1 | 4400 V, 0.5 mA | I | | 4.0 ± 1.8 | 0.18 ± 0.41 | N/A |
| Zener Diode | | | | | | | | | |
| | Parallel Array (ZM4728AGS08) | | d = 2.6 mm, l = 5.0 mm, N = 1329 | 2.4 V, 2 A | I | | 3.7 ± 2.9 | 0.3± 1.1 | N/A |

* PZT Material Navy Type VI, ** calculated from measured capacity, *** below equivalent photon pressure force < 44 nN

C-axis = coil central axis, d and t .. capacitor/varistor diameter and thickness, $d_o/d_i$ .. coil outer and inner diameter, h .. coil height, N .. coil number of turns ($N_t$ … toroid turns, $N_c$ .. coil turns for crossed-coil)

Table 1 Experimental Result (All Material Constants are Typical Datasheet Values Unless Stated Otherwise)